# RAMANMETRIX: a delightful way to analyze Raman spectra

Darina Storozhuk[1,2 #], Oleg Ryabchykov[1,2 #], Juergen Popp[1,2 *], Thomas Bocklitz[1,2 *]

[1]Leibniz Institute of Photonic Technology Jena, Albert-Einstein-Str. 9, 07745 Jena, Germany,

[2]Institute for Physical Chemistry and Abbe Center of Photonics, Friedrich Schiller University, Helmholtzweg 4, 07743 Jena, Germany.

[#] equal contribution
[*] corresponding authors

## Abstract

Although Raman spectroscopy is widely used for the investigation of biomedical samples and has a high potential for use in clinical applications, it is not common in clinical routines. One of the factors that obstruct the integration of Raman spectroscopic tools into clinical routines is the complexity of the data processing workflow. Software tools that simplify spectroscopic data handling may facilitate such integration by familiarizing clinical experts with the advantages of Raman spectroscopy.

Here, **RAMAN**METRIX is introduced as a user-friendly software with an intuitive web-based graphical user interface (GUI) that incorporates a complete workflow for chemometric analysis of Raman spectra, from raw data pretreatment to a robust validation of machine learning models. The software can be used both for model training and for the application of the pretrained models onto new data sets. Users have full control of the parameters during model training, but the testing data flow is frozen and does not require additional user input. **RAMAN**METRIX is available in two versions: as standalone software and web application. Due to the modern software architecture, the computational backend part can be executed separately from the GUI and accessed through an application programming interface (API) for applying a preconstructed model to the measured data. This opens up possibilities for using the software as a data processing backend for the measurement devices in real-time.

The models preconstructed by more experienced users can be exported and reused for easy one-click data preprocessing and prediction, which requires minimal interaction between the user and the software. The results of such prediction and graphical outputs of the different data processing steps can be exported and saved.

# 1 Introduction

Raman spectroscopy is a vibrational spectroscopic technique that showed its potential for biomedical and clinical applications. As a label-free non-destructive technique, it requires minimal sample preparation of biomedical samples[1] and can, potentially, be applied *in-vivo*[2,3]. Despite the numerous benefits, Raman spectroscopy is not widely used in clinics because the raw spectra typically cannot be directly interpreted. To convert the spectral data into easy-interpretable information, data processing workflows, including data preprocessing and chemometric methods, are needed[4]. Automation and optimization of such workflows have a very high barrier of entry and require expert knowledge on the application field itself, Raman spectroscopy, programming, and statistics.

The usage of biomedical Raman spectroscopy can be simplified by the introduction of user-friendly software tools specialized in this specific type of data. Such software can increase the interest of clinicians in Raman spectroscopy, which can widen this research field and increase the presence of Raman spectroscopy in biomedical applications.

## 1.1 Raman spectroscopy in biomedical and clinical applications

Vibrational spectroscopy techniques, such as IR and Raman spectroscopy, provide a good overview of the chemical composition of different kind of samples. These techniques have a less complicated implementation than mass spectrometry and give deeper insights than label-free fluorescence spectroscopy. For the analysis of biological samples, Raman spectroscopy is more convenient than IR because biological samples typically contain water, which features a significant IR absorption that overlays with all other peaks. Due to the differences in selection rules between IR absorption and inelastic Raman scattering, there is no issue with the water peaks in Raman spectroscopy, which makes it suitable for non-dried samples[5].

Due to these advantages, Raman spectroscopy can be used for multiple biomedical applications. For example, Raman spectroscopic identification of bacteria is possible without time-consuming cultivation processes[6]. This makes it a lot faster than the conventional techniques[7]. Raman spectroscopy can detect pathogenic bacteria[8] and differentiate between fungal spores[9]. The testing for antimicrobial resistance might be performed on a single-cell bacteria level[10]. Besides pathogen identification, Raman spectroscopic techniques can be used in diagnostics[11] of infection and sepsis[12], kidney disorder[13], breast cancer[14], colon cancer[15], hepatocellular carcinoma[16], brain tumor[17], etc. Bringing Raman spectroscopy into the clinical routine can speed up diagnostics, and therefore, improve patient outcomes.

## 1.2  Corrupting effects in Raman spectroscopic data

To use Raman spectroscopy and its respective data for biomedical tasks stated above, the data needs to be corrected for artifacts and standardized. A main reason for the correction is that Raman spectroscopic data can be corrupted by numerous effects which can dramatically influence the analysis outcome. Some of the effects are related to the samples and the measurement process, while others are non-sample related and originate from the device or external effects. In the context of this article, we will refer to the non-sample related effects as extrinsic effects.

The main extrinsic corrupting effects are cosmic ray spikes and improper device calibration. The spikes are originating from high-energy cosmic particles that hit the CCD camera of the spectrometer. The spikes appear in spectra as sharp narrow artifacts with high intensity[18]. Such artifacts are also referred to as outliers because they appear at random positions within the spectrum. Another extrinsic issue is the device calibration, however, this can usually be ignored if the analyzed data are collected within a single set of measurements from the same device, but in real-life applications, a proper device calibration is of utmost importance[19]. The intensity response, dark current of the detector, and wavenumber calibration may significantly vary between devices and even change over time for the same device, which can corrupt the data analysis. For suppressing the extrinsic corrupting effects, correction procedures should be applied. In this article, we will refer to the correction procedures for the points above as data pretreatment (see 5.1 Data pretreatment).

Besides the extrinsic effects, a significant corrupting contribution to the raw Raman spectra of biological samples is based on the sample's fluorescence. A good practice is to select the measurement parameters in a way that minimizes the contribution of the fluorescence. However, avoiding the fluorescence is rarely possible, so additional baseline correction is required to suppress the influence of the fluorescence background on the analysis[20]. Along with the fluorescence contributions, the offset of the focus or sample inhomogeneity can impact the intensity of the measured spectra, which leads to variations in absolute intensities between measurements. As a result, the intensities are not directly comparable and require normalization. For further details on the baseline correction and normalization procedures, see subsection 5.2 Data Preprocessing.

*1.3 Related work*

Most of the producers of commercial Raman spectroscopic devices provide software that is capable of essential spectral preprocessing and analysis. Unfortunately, these software tools are device-specific and do not reach the level of robustness that is needed for clinical applications. Besides the device-specific software, there are also programs that are focused on the specific type of analysis, such as spectroscopic imaging or mapping (Vespucci[21] and RamanToolSet[22]).

One of the best examples of the program that makes it possible to build, validate, and apply supervised models is Raman Processing Program[23]. Another example is Spectragryph – optical spectroscopy software[24], which has a huge number of various optical spectroscopic data visualization and analysis features but might also be overwhelming for users that need a more specific set of features. Unfortunately, even these programs do not cover sufficient steps of wavenumber and intensity calibration or advanced cross-validation options, which makes them not suitable for constructing robust Raman spectroscopic data analysis pipelines for predictive modeling in a clinical setting.

Currently, researchers routinely use different MATLAB, R, and Python scripts and respective libraries for Raman spectroscopic data processing and analysis. These scripts can perform very well and have great flexibility in usage. The main drawback of such an approach is that the researcher needs decent skills and experience in a particular programming language and its libraries, as well as a good understanding of the specifics of Raman spectroscopic data processing and some experience with machine learning.

To our knowledge, **RAMAN**METRIX is the first software with a user-friendly graphical user interface (GUI) that provides the complete workflow for Raman data pretreatment, preprocessing, quality assessment, robust cross-validation of supervised models, and application of the models to the independent test data. Furthermore, it does not require any additional packages and can be installed as a stand-alone application or can be accessed online using a web browser.

## 2 Application architecture

**RAMAN**METRIX is based on a divided backend-frontend concept for the software architecture. Here, the client-side (frontend) and the server-side (backend) of the program communicate with each other through an application programming interface (API). The API is implemented using HTTP-based representational state transfer (REST) architecture, which makes it possible to communicate with the backend not only through the user interface, but also from other applications. The client-side is represented by a web-based single page application (SPA) with a component-based structure, written in Vue.js, a progressive JavaScript framework. For the frontend design development, we used HTML5, CSS, Bootstrap 4, and JavaScript with various open-source JavaScript packages. The packages were managed using the npm package manager for Node.js. Data visualization elements were programmed based on dygraphs[25] and D3.js[26] JavaScript libraries. Axios[27] module was utilized to send synchronous and asynchronous HTTP requests to the server-side in a JSON format. Choosing web technologies for the GUI has several advantages: it has a responsive layout for compatibility with different screen sizes, it has a consumer-attractive-looking design, and it gives a possibility to create a Desktop/Web hybrid application. To keep the graphs and user information, the SPA application uses the local storage and the Indexed Database API of the web browser.

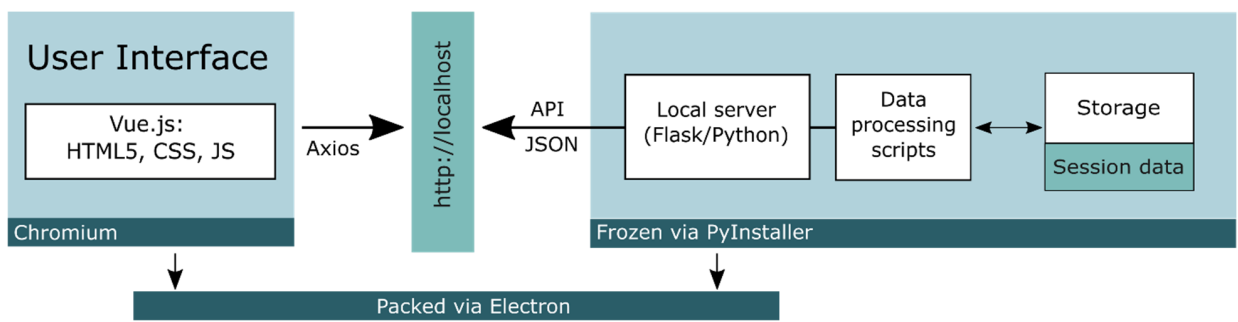

Figure 1 The architecture of the desktop version of **RAMAN**METRIX. The left part of this scheme shows frontend application which is running by Electron's Chromium browser engine. The right side shows which parts of the backend system is wrapped via PyInstaller. Frontend and backend parts are communicating with each other via REST APIs.

The backend side was built in Python and it is accessed by SPA using the endpoints of API. The backend manages session data and performs all the heavy computational tasks such as spectral data processing, and machine learning. The data processing scripts were written using scientific computing Python packages as NumPy[28], Pandas[29], scikit-learn[30], SciPy[31], TensorFlow [32].

## 2.1 Desktop version

To provide a desktop version of the software, the above-described system components (Figure 1) are wrapped using Electron[33] - the framework for building a cross-platform standalone software that uses the Chromium browser engine. It allows us to use web technologies for GUI development. The SPA application was compiled into a static file, executable by the Chromium browser.

To keep the application lightweight, the Flask[34] micro-framework was used for the backend. Fast access and simple storage of session data was implemented using the session storage, based on SQLite[35] and sqlite3[36] python package. To handle the backend code, the Python data processing scripts with the required dependencies were assembled by PyInstaller[37] into an executable program.

The SPA and the executable backend were wrapped together by the Electron framework into a single installation file. The current version is packed for a 64-bit Windows 10 system. Although, support of other operating systems is theoretically possible with such architecture, it was not implemented due to the lack of demand. The important advantage of such architecture is that the backend can be directly accessed from other programs through HTTP requests, which may be helpful for automated tasks, such as process control. The backend can even be executed separately, without GUI, by launching the "ramanmetrix_backend.exe" executable.

## 2.2 Online platform

The flexible frontend-backend architecture of the software made it possible to deploy **RAMAN**METRIX online with minimal modifications (see Figure 2). The online version is accessible from the internet and demands an additional authorization interface for parallel multi-user access. The multi-user access requires handling and storage of user information and user sessions, including storage of large amounts of temporary data (e.g. spectra after each preprocessing step) for each user. To fulfill these requirements, Django[38] framework was chosen for the online platform instead of Flask microframework, which is used in the desktop version.

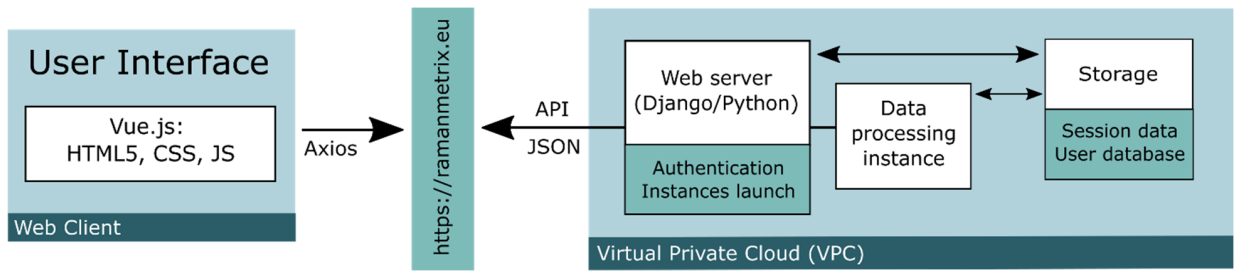

Figure 2 The schematic architecture of the **RAMAN**METRIX's online version. The web server can be accessed online and used for authorization and launching computational instances. The computational instance itself is not accessible from outside of virtual private network. The web server is used as a proxy between the user and computational instance to avoid unauthorized access.

Currently, the online platform is set up at Amazon Web Services (AWS), where authorized users can launch elastic compute cloud (EC2) instances with a necessary amount of memory (see Figure S1). After the instance is initialized, the **RAMAN**METRIX interface, identical to the desktop version, opens in the browser.

Besides EC2 service, the online platform uses other services for storing authentication and session data in a relational database service (RDS), large temporary session objects in the simple storage service (S3). The storage of sessions and temporary data preserves the analysis results even if the instance was terminated. Users can only access the instances that they launched themselves to avoid unexpected limitations in computational power due to the shared usage of the instance. It is possible for a single user to run up to 2 instances concurrently, but only one project per session can be opened in parallel. For running independent analyses in parallel, different user sessions should be established for each project. In order to establish multiple user sessions, different approaches can be used, such as logging in from different devices or browsers, as well as by using multiple browser profiles or an incognito browser window.

## 3   User experience

The desktop version of the program can be started as any other windows application, e.g. by a double clicking on the **RAMAN**METRIX icon either on the desktop or in the start menu. When started for the first time, the window for submitting a license will be displayed to the user. The window of the program with a valid license is presented in Figure 3. The interface is divided into three zones: the top navigation bar, the left menu bar, and the main content area. Through the universal top navigation bar - stepper, the user can get to any data processing step. For greater convenience for users, the completed and the current data processing routine steps are highlighted in different ways.

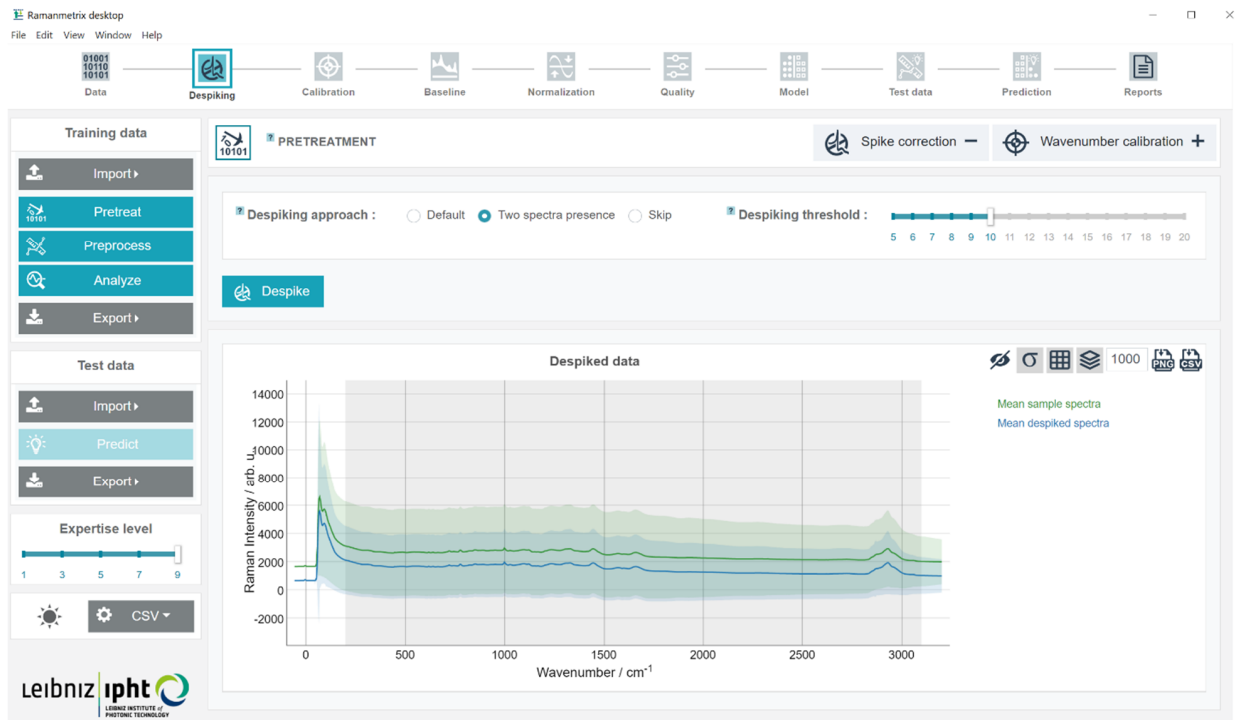

Figure 3 Snapshot of the Rammanmetrix desktop window. This view corresponds to the moment when the training dataset has been imported and the despiking step is completed. The border around the icon in the stepper shows which step is currently visible for the user ("Despiking"). Additionally, the current step (that is in progress or the last step that has been completed) is highlighted with a turquoise color. In this screenshot, the mean spectra are plotted with a manually set y-offset of 1000 arb. u., and the gray area indicates the wavenumber area that will be used in the next step of the data processing. The interactive plot makes it possible to investigate the peak positions/intensities and zoom in into different spectral areas.

The left menu is divided into two panels to perform actions related to the training or test datasets. Both panels have "Import" and "Export" dropdowns with different types of options for the export and import of data (see section 6 Exporting models and results) as well as buttons for executing the calculations. The full preprocessing and model training pipeline can be executed with a single click on the "Analyze" button. For more advanced usage, there are buttons to execute data pretreatment and data preprocessing procedures one by one. Additionally, there is an execution button for each step of data processing. Under the "Test data" panel, there is a slider to change the user "expertise level". This feature makes GUI customizable depending on the user's level of knowledge in the field of processing and analysis of Raman spectra. The higher level is chosen, the more parameters the user will be able to adjust. The rest of the parameters will be simply hidden so that UI will not be too overwhelming when there is no need for it.

Below the "expertise level" slider, there is an icon to switch between dark and light mode. The graphic content can also be saved in both themes. By clicking on the "CSV" button, the user will have the option to select the decimal separator and column delimiter symbols for the exported CSV files.

At the top of the main content area, an additional minor navigation panel is used to group data processing steps and control their visibility. The screenshot in Figure 3 illustrates that the "Spike correction" tab is open and the "Wavenumber calibration" tab is collapsed. Both tabs, furthermore, belong to a "PRETREATMENT" section.

Each block of the training data workflow is structured in a similar manner. The block starts with a panel for the adjustable parameters and buttons to start the calculation. For the completed steps, there is also a graphic visualization of the results at the end of the block. If the user wants to perform calculations for a single step, but the parameters in some of the previous steps changed, a dialog box appears where one can decide whether to recalculate the previous steps or discard the changes.

To facilitate the work with the program, e.g., when large number of parameters need to be adjusted and the user is on a high expertise level, some reference information can be obtained directly from the application. To open a text tooltip, the user needs to hover over the question mark icon next to the parameter name. A similar tooltip is available to the left of the title ("PRETREATMENT" in Figure 3) on the minor navigation panel. If some errors occur in the computational backend, the user will see a message box with an error description in the GUI. The process of data import, as well as active calculations, is accompanied by an animation and these processes can be terminated by a stop button.

The results are visualized by interactive plots. The user can zoom in on a specific range of values along a horizontal or a vertical axis. On the plots with multiple spectra, the user can highlight a spectrum and a respective legend by hovering the mouse over the spectrum or the legend. The highlight can be also locked. Checkboxes in the legend enable the selection of the visible spectra in the plots. Other features can be enabled using buttons on the graph's toolbar, above the legend. For better visualization of graph series, the user can set a custom offset value on the y-axis. In addition, there are visibility switches for the standard deviation and grid lines. Graphs accessible from despiking and baseline correction steps also visualize the selected wavenumber range. When the user changes the range in the calibration block, the gray background area on the "Despiked data" plot will be reactively changed. The far-left button in the toolbar can make this area invisible on the chart.

# 4 Data input

**RAMAN**METRIX supports Raman spectroscopic data of different formats including TXT tables and SPC file format. Full information about currently supported data formats and devices is available at [https://docs.ramanmetrix.eu](https://docs.ramanmetrix.eu). Prior to the import of the spectra to **RAMAN**METRIX, the data files should be compressed into a single ZIP file. It is expected that some information about the measured samples and the measurements should be provided along with the data. In the following text, such information will be referred to as metadata.

Providing metadata is important for the correct identification of labels, measurement dates, and other information linked to the imported spectra. For simple tasks, the metadata can be parsed from the folder structure automatically, but in most cases, metadata should be provided as an additional table. The metadata table can be provided as CSV files (*.csv), or MS Excel files (*.xlsx; *.xls) with a mandatory "metadata" word in the filename. The metadata table should be placed inside the ZIP file (*.zip) together with the spectral data. Alternatively, metadata can be also imported separately, directly after the spectral data import.

The list of default column names (case sensitive) in the metadata table can be found at [https://docs.ramanmetrix.eu](https://docs.ramanmetrix.eu). Any other custom columns may be added to the metadata to be used as an alternative class or batch labels. The addition of custom columns with numeric values is essential to supply numeric responses for the regression models.

Although using the folder structure instead of the metadata table may be simpler for a quick data overview, the experience shows that keeping the folder structure correct for moderate and large data sets can become quite challenging. To avoid possible confusion related to generating a suitable folder structure, it is highly recommended to import data along with the metadata table.

# 5   Data processing pipeline

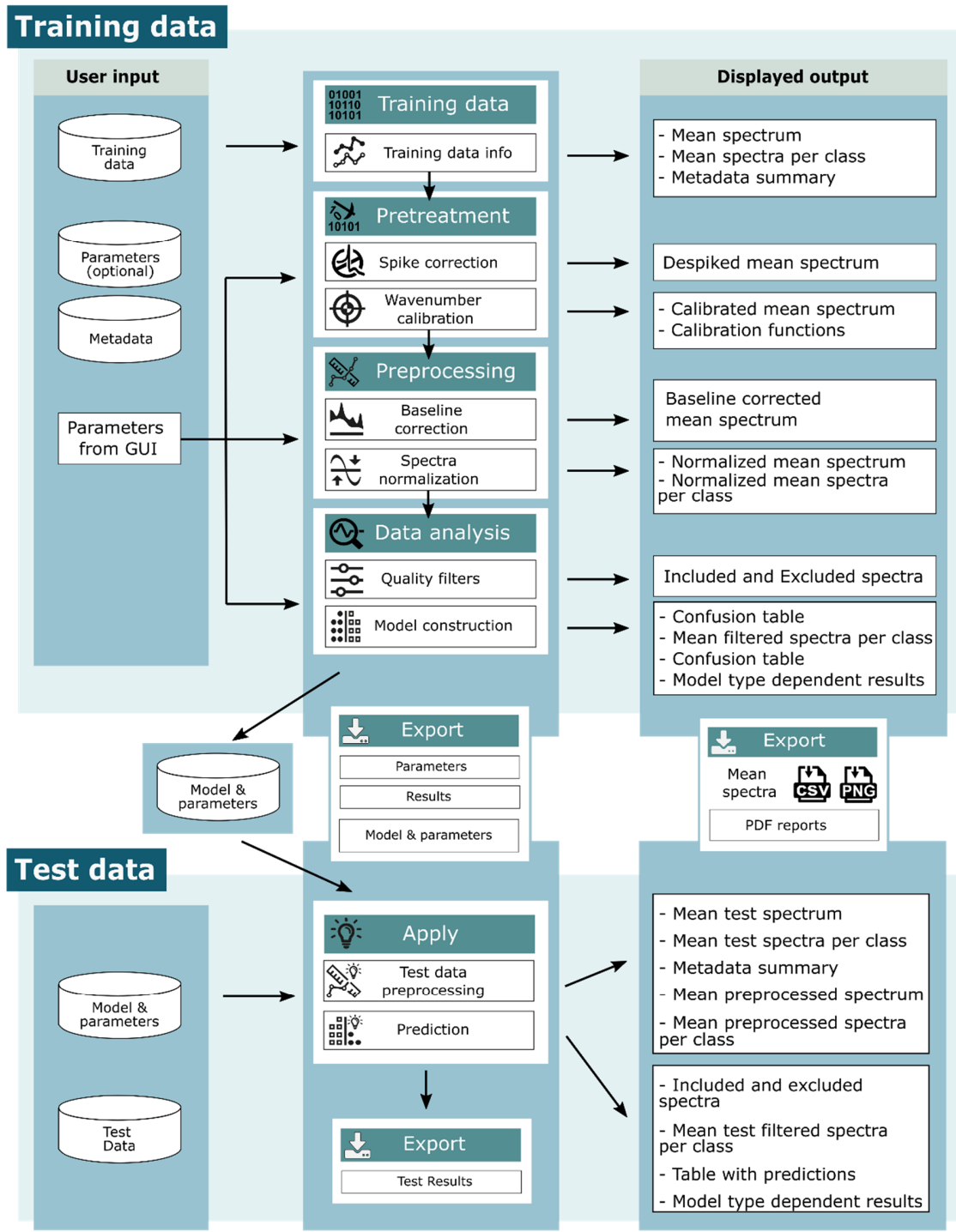

Figure 4 Flowchart of the **RAMAN**METRIX GUI. The data processing flow is divided into two parts: training and testing. The parameters in the training phase can be adjusted through the GUI. In the test phase, the parameters and the model cannot be changed. The constructed model can be used for testing straight away or exported, and then the saved model can be reused for testing at any time.

The data processing workflow implemented in the **RAMAN**METRIX GUI is illustrated in Figure 4. In the left column, the possible user input data are shown. In the training data workflow, parameters and metadata can be imported separately from the spectral data to re-use parameter settings. In the middle column, a sequence of data processing steps, and how they are grouped is presented. Each step is described in detail within the following subsections. The model from the training workflow can be directly used in the test part of the software. If the lowest expertise level is used, only the test data workflow will be visible, and the preconstructed model needs to be imported for prediction. In the right column, the list of the graphical results for both workflows is shown.

In order to demonstrate the capabilities of the GUI, a Raman dataset from [39] was used. The dataset consists of six bacterial species for which the Raman spectra of the nine biological replicates (batches) were cultivated. It was divided into training dataset (first seven biological replicates) and test dataset (replicates eight and nine). Each biological replicate was considered as an independent batch in the training data. The metadata file for the whole dataset can be downloaded from https://github.com/Bocklitz-Lab/supplementary_files/. As soon as the data is imported, the user can get a first overview of the raw training data prior to the data processing. The first step available in the top navigation menu, "Data", contains two tabs with raw mean spectra and metadata information. In the first tab, there is a mean spectrum over of the raw training data and mean spectra per class (classes names are taken from the "type" column in the metadata file). In the second tab, the information from the provided and automatically detected metadata is summarized in lists. In this summary, users can check whether the software determined the measurement dates correctly (in case they were not indicated in the "date" column of the metadata file), since this is critical for the calibration step. A separate list shows for which measurement dates the standard spectra are available.

## 5.1 Data pretreatment

Pretreatment steps are meant to suppress extrinsic corrupting effects such as cosmic ray noise and variations in the device calibration. The despiking algorithm detects and removes sharp features with high intensity. These features are called spikes and they are generated by high-energy cosmic particles that hit the CCD camera. The default approach is a recommended option suitable for most cases. It uses a modification of the spike detection algorithm [16] with one-dimensional Laplacian and a threshold which can be manually adjusted in the GUI. Prior to applying the threshold for spike detection, the Laplacian response of each spectrum is centered to the median smoothed value of the response and then normalized to the robust standard deviation. In the default option, the detected spikes are replaced by the values of median smoothed spectrum.

In the case when two spectra per point were measured, "Two spectra presence" despiking option should be chosen to use pairs of spectra for despiking and obtain an averaged corrected spectrum from each pair. When two spectra are used, the threshold is applied to the normalized absolute difference between the Laplacian responses of these spectra. After the spike localization, it is replaced by the sum of the original spectrum's baseline and the baseline corrected values from the second spectrum in the spectral pair. The despiking step can be also skipped when, for example, the procedure has already been performed by the measurement device itself.

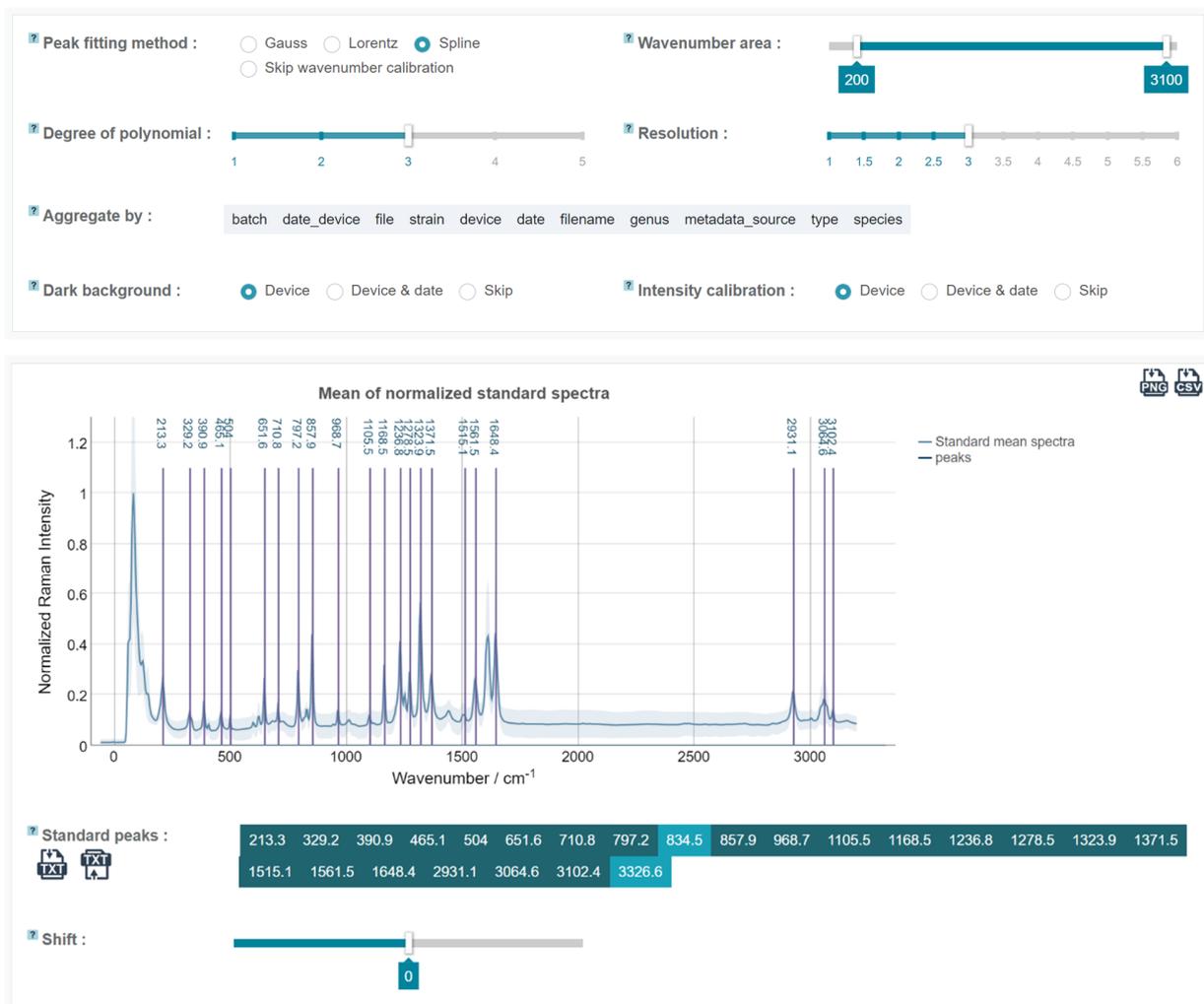

Figure 5 Calibration parameter block. The average measured standard spectrum (here 4-acetamidophenol) is plotted along with the tabled positions of the reference peaks. If different calibration standards should be used, a list of tabled peaks should be imported as a TXT file. The peaks can be added/removed with the buttons below the plot. In case the data has a systematic shift in calibration, it can be specified manually to initially minimize the shift between the tabled and the detected peaks.

The calibration step consists of wavenumber, dark background, and intensity correction steps. The wavenumber calibration approximates all the sample spectra onto the same wavenumber axis, using calibration functions built based on the peak positions in a standard spectrum. The calibration is performed based on the positions of peaks within the spectrum of a wavenumber standard substance (4-acetamidophenol) [19]. Thus, information on the presence of standard spectra should be provided in the metadata file. To perform wavenumber calibration, peak positions should be detected. For that purpose, the peak is approximated with one of the following shapes: Gaussian, Lorentzian, or cubic Spline. The polynomial degree of the calibration function can be chosen by the user (Figure 5). The skip option here means that standard spectra are ignored, but all spectra will be approximated to a new linear wavenumber axis. The wavenumber range and the resolution for the new wavenumber axis can be adjusted with respective sliders. Spectra can be averaged based on the selected grouping label. In addition to the above configurations, the user can examine the mean standard spectrum (Figure 5) and manually select standard peaks. It is recommended to select only clearly visible peaks for wavenumber calibration. The tabled positions of the peaks are taken from McCreery[40]. Users can also import another file with standard peaks if a different material was taken as a standard substance. If the peak positions are too far from the tabled positions, the wavenumber axis can be manually shifted. To evaluate the calibration results, in addition to the mean calibrated spectra, the difference between the original and calibrated wavenumber axis is plotted (Figure S2) for each measurement date.

If data measured on different devices are analyzed together, intensity calibration is paramount. The intensity calibration requires the intensity reference and the measured intensity uploaded to the software. The reference should be imported along with other parameters or wavenumber calibration peaks as "calib_response" parameter. The reference can be provided as the intensity values with the respective wavenumber positions or described by a set of polynomial coefficients. The measured intensity standard must be imported along with the data and marked in the metadata table with a "True" Boolean value in the column "standard_intensity" (for the rest of the data "False" value has to be set). If the measured standard is available and specified in the metadata, the user can choose whether to apply intensity calibration for each device and each date or only for each device.

Dark current background (the signal with the shutter closed) subtraction from the spectrum is an optional data pretreatment step, which may be skipped if no intensity calibration is applied and the baseline correction is applied later, but it is needed for a proper intensity calibration. For subtracting a measured background, it must be imported along with the data and marked in the metadata table with "True" value in the column "dark_bg" (for the rest of the data the value "False" must be set). Dark background correction, similar to intensity calibration, can be applied per device and date (same as wavenumber calibration), or per device only.

## 5.2 Data Preprocessing

Data preprocessing steps standardize the data by suppressing variations in the fluorescence background and variations in the intensity of the signal related to focus fluctuations. Removal of the fluorescence background is one of the main objectives in the Raman data preprocessing pipeline for biological samples.

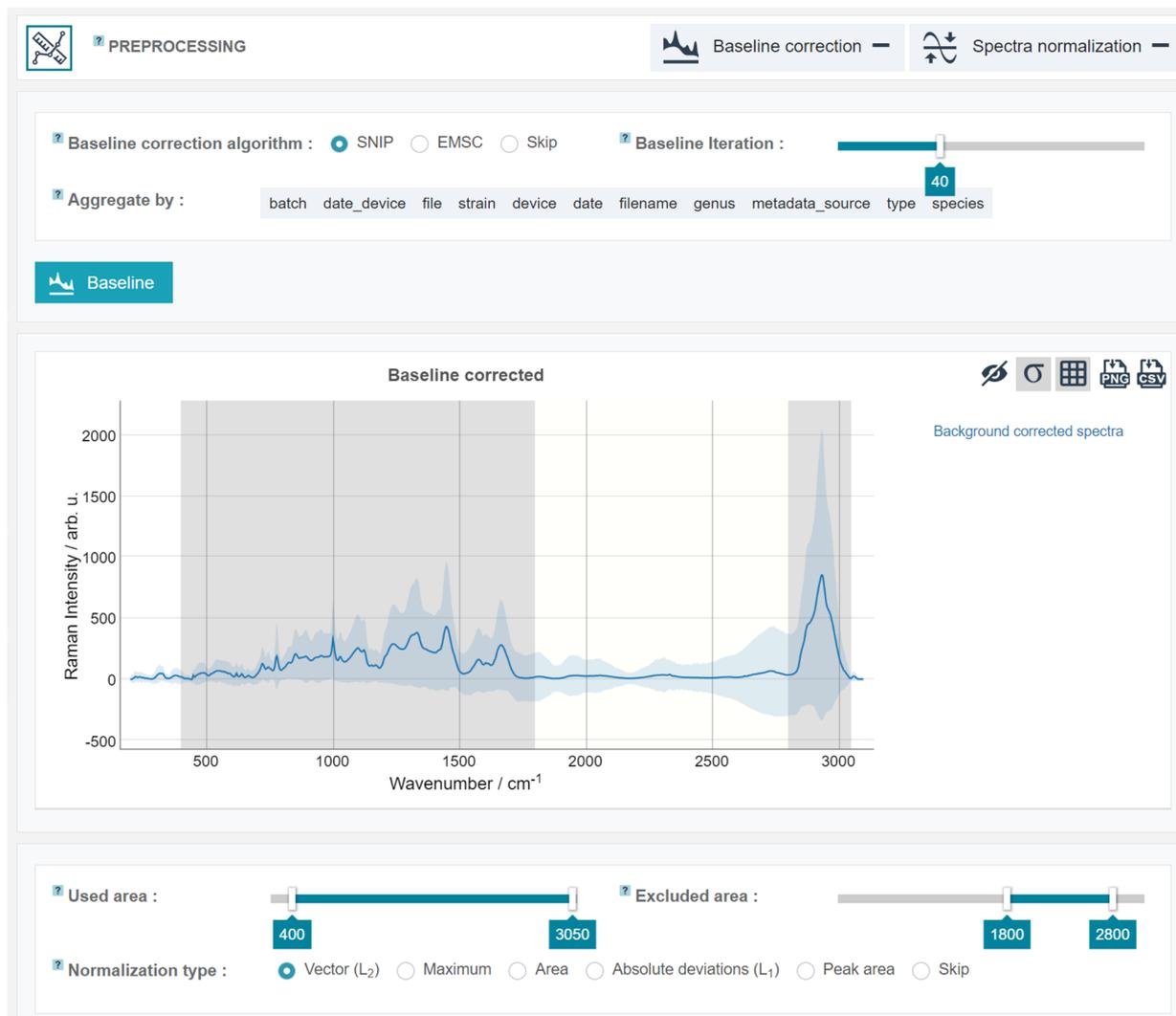

Figure 6 Screenshot of the preprocessing steps. It shows the set of parameters of the calibration tab, the resulting baseline corrected mean spectrum, and the parameters block for the normalization step. The baseline corrected spectrum is utilized on the normalization step to vividly illustrate the used and excluded wavenumber area using colored regions.

Baseline correction estimates the fluorescence background mathematically. If performed well, baseline correction may dramatically improve the outcome of the analysis step. Two baseline correction algorithms are available in **RAMAN**METRIX (Figure 6). The first algorithm, the sensitive nonlinear iterative peak (SNIP) clipping algorithm [41–43] is a robust iterative background estimation method. To decrease the influence of noise in the spectra, smoothing is applied at the first iteration of the baseline estimation. The number of iterations can be set manually by the user. If the number of iterations is too low the background is overestimated, while large number of iterations underestimate it. A higher number of iterations should be preferred for the spectra with wide peaks and high resolution. The second algorithm is a model-based preprocessing using the extended multiplicative signal correction (EMSC) [44]. This method preprocesses spectra to make them similar to the reference spectrum. If no spectra are marked in the "reference_sample" column of metadata table, then a mean over the training dataset is used as a reference. The polynomial degree of the baseline can be specified through the GUI. When EMSC is applied, normalization is unnecessary and may be skipped. As with the calibration step, the user can group and then average the baseline-corrected spectra.

The last preprocessing step is the normalization of Raman spectra, which decreases the variations in the data caused by imperfect focusing while acquiring the spectra. Various normalization approaches may be selected: vector ($l_2$), to the maximum intensity within the spectrum, to the integrated area, to integrated absolute intensities ($l_1$), to the integrated peak area. The user can also choose the range of wavenumbers that will be included in further analysis and add a "silent area" range, that will be excluded. The "silent area" is specific to biomedical samples as no Raman peaks are expected within the range 1800-2800 cm$^{-1}$. Spectral data can be cropped using these options even if the user chooses to skip normalization. The results at the normalization step are presented by the mean normalized spectrum over the entire dataset and the mean spectra for each type (e.g., bacteria species in case of the example data).

### 5.3 Quality check

To improve the analysis efficiency, the user may filter out low quality spectra (only available for the high expertise levels). By default, all filters are disabled. To activate the filter, one can click on the checkbox or directly on the boxplot. In the active filters, a separator appears that allows changing of the filtering threshold. The threshold value is displayed directly above the boxplot. At the time of writing, there are eight filters available, and they are shown in the Figure S3.

The first pair of filters allows filtering data based on the minimum and maximum integrated normalized intensity values within a certain wavenumber range. The quality peak range can be specified using a slider (only available when the filter is active) and visualized on the normalized mean spectra plot above the filters. After changing the range of wavenumbers, the user should click on "Update quality peak" and thereafter set the threshold for these filters. The next pair of filters utilized correlation with a pretreated (after the calibration) or preprocessed (after the normalization) reference spectrum. The reference spectrum should be defined by the "reference_sample" column of the metadata table, otherwise the mean over the training dataset is used. Providing the reference for the correlation-based filters is especially important when a large number of outliers is expected. The threshold for the next filter can be set as a minimum value of the signal-to-noise ratio (ratio between maximal intensity and a standard deviation of noise). The signal is estimated as a mean of a smoothed spectrum and the noise is estimated as a standard deviation of a difference between an original and a smoothed spectrum. In the next filter, the maximum value of the integrated intensity of the estimated baseline is used as a parameter. In the case of model-based EMSC preprocessing - an intercept model coefficient is used as the integrated background intensity. If the baseline correction is skipped, the minimum value within the spectrum is used instead. The last pair of filters allows filtering of the data by the minimum and maximum value for the integrated intensity of the baseline corrected spectrum (values calculated before the normalization).

After applying the filters, the mean spectra of the data that passed or failed the quality control will be displayed. Above the filters, it will be shown how many spectra from the dataset passed the quality check.

## 5.4 Analysis

There are three groups of analysis methods available in **RAMAN**METRIX: classification, regression, and clustering. The following supervised classification methods can be utilized: linear discriminant analysis (LDA), support vector machine (SVM) classification, random forest (RF), *k*-nearest neighbors (*k*NN), and two different implementations of partial least squares discriminant analysis (PLSW2A-DA and PLS2-DA). If purely numerical responses are available, the regression models can be applied: linear regression model (LM), SVM, and two algorithms for partial least squares regression (PLSW2A and PLS1). Besides the supervised methods, an unsupervised Ward hierarchical cluster analysis (HCA) with a selected number of clusters can be utilized to obtain an overview of the data.

In order to avoid overfitting, dimensionality reduction is used prior to the analysis for the majority of the methods. Only PLS-based methods are used without dimensionality reduction because PLS is a supervised dimensionality reduction method itself. The PLS implementation that uses Wold's two-block, mode A PLS (PLS-W2A) [45] estimates using the singular value decomposition (SVD) algorithm. Both PLS1 and PLS2-DA methods use NIPALS algorithm.

Figure 7 Selection of the possible model type and validation strategy. With the parameters selected in the screenshot, the user constructs a PCA-LDA model to classify the bacteria according to the labels specified in the "type" column in the metadata table. The number of features will be optimized automatically, and the model will be evaluated using the 2-level leave-batch-out cross-validation (CV) method. Switching to a 1-level can be done by setting inner evaluation type into "None".

For non-PLS-based methods, there are 2 unsupervised dimensionality reduction approaches available. The first method for dimensionality reduction is based on principal component analysis (PCA) and the other one is based on convolutional neural network (CNN) feature extraction. At the time of writing the article, the CNN-based dimensionality reduction is only available for LDA, LM, and HCA. This approach is based on the feature extraction part of an architecture published by Liu et al. [46]. Subsequently, the extracted features are reduced in dimensionality either by PCA or by the autoencoder with a selected number of the bottleneck features. To use the autoencoder for the dimensionality reduction, the number of epochs should be set to a non-zero value. If the number of epochs is set to zero, the dimensionality reduction of CNN features is performed by PCA. The full list of models with the respective dimensionality reduction approaches is presented in Figure 7.

The supervised models can be validated by a 10-fold or a leave-batch-out cross-validation approach. For larger flexibility, the class labels and batch labels can be selected from the columns of the provided metadata table. Either single columns or combined labels, based on multiple columns, can be used as customized labels for cross-validation. The "Included labels" parameter allows the user to exclude spectra corresponding to any value of any of the metadata columns. And only those spectra that will satisfy all the conditions will be used further for model construction. Thus, the user does not need to restructure the data and metadata and then import it again but can simply exclude a part of the data before constructing a model. For the regression, any of the numeric columns in the metadata can be selected as a response, so a regressions with different responses can be constructed without reloading the data and repeating the preprocessing steps.

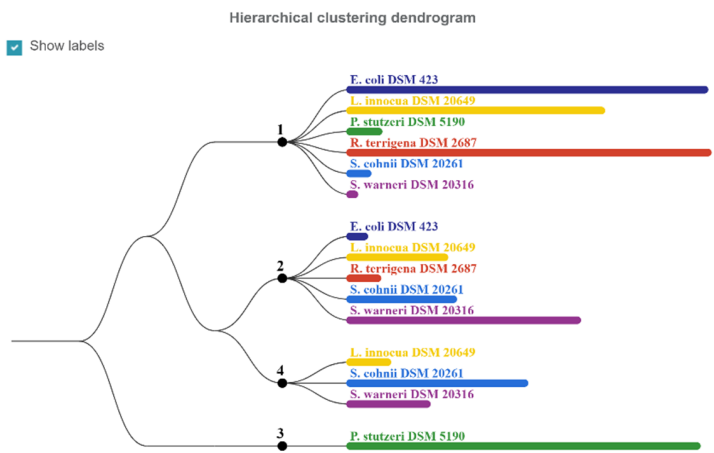

Figure 8 Cluster dendrogram example for four clusters. The black lines of the dendrogram represent the hierarchy of clusters. The bars are colored and labeled according to the information provided in the metadata table. For a more compact view of the dendrogram, the labels can be hidden by unclicking the "Show labels" checkbox.

The output results shown to users differ largely on the selected analysis type and model. The output for classification models contains a confusion table and classification metrics for cross-validated predictions. The regression model output consists of a table with metrics for cross-validated predictions and a regression plot. The metrics are not available for the clustering model, as this is not a supervised method. The clustering results are visualized using a confusion table and a cluster dendrogram as illustrated in Figure 8.

Some supervised models are visualized with a scatterplot with fitted values as presented in Figure 9. In case of clustering or supervised models that do not support scatterplot visualization (RF and *k*NN classification, LM and SVM regression), a scatterplot for an unsupervised feature extraction (e.g. PCA) is generated. Along with the scatterplots of LDA, PLS, or PCA scores, model loadings and mean spectra per class are also shown at the output of the model step (Figure 10).

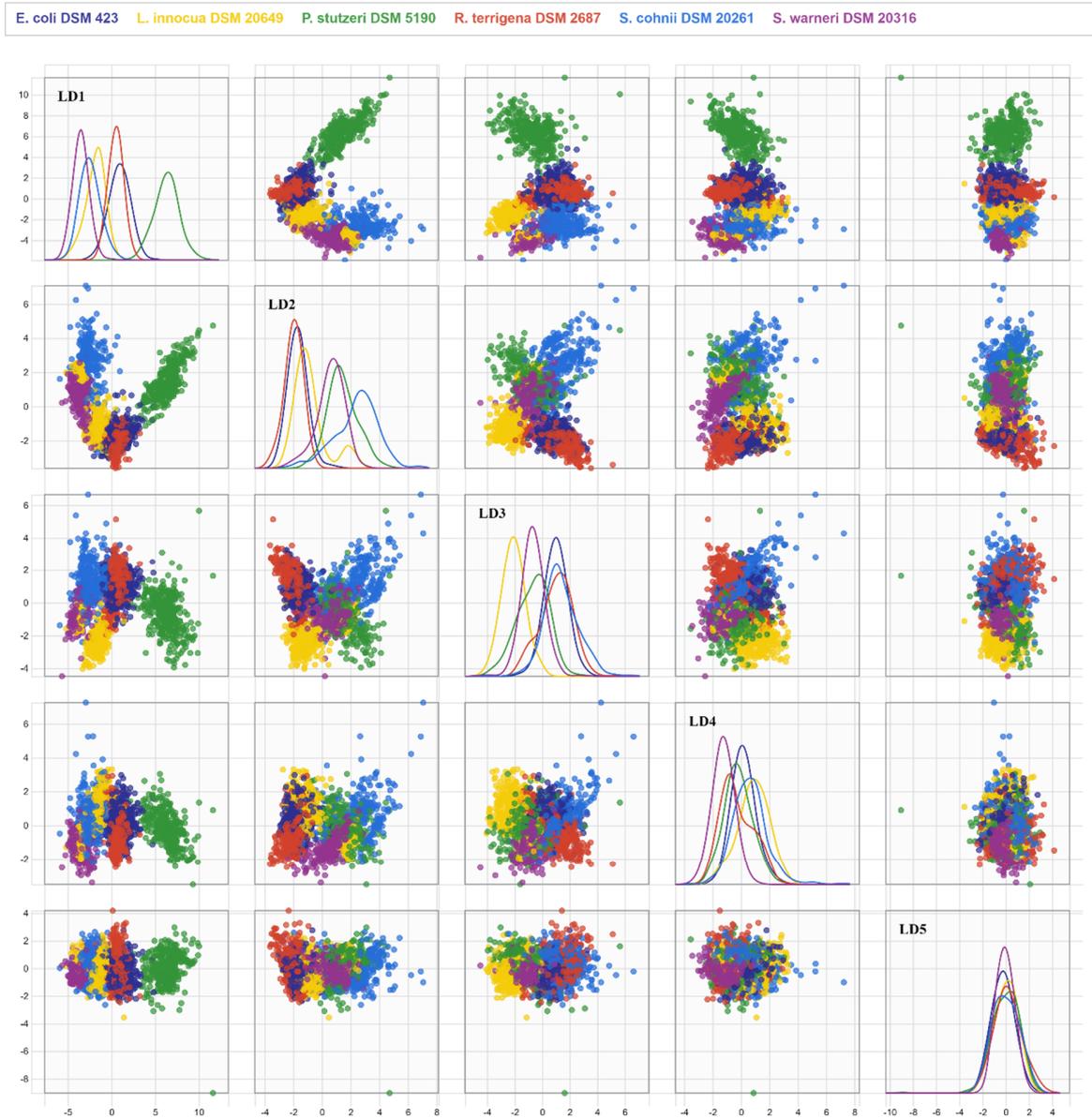

Figure 9 Scatterplot example for PCA-LDA model on bacteria species level. This scatter plot visualizes the classification of the bacterial species: it shows, for example, that *P. stutzeri* (green) is well separated from the other species, and *R. terrigena* (red) and *E. coli* (dark blue) are highly overlapped.

As the number of features (e.g., principal components or PLS latent variables) used in the model is a very important parameter, it can be optimized automatically within the cross-validation routine. For obtaining a better estimation of the model performance, users can apply a 2-level cross-validation by enabling the inner cross-validation loop[47]. This kind of validation makes it possible to estimate not only the model itself but also the performance of the optimization routine.

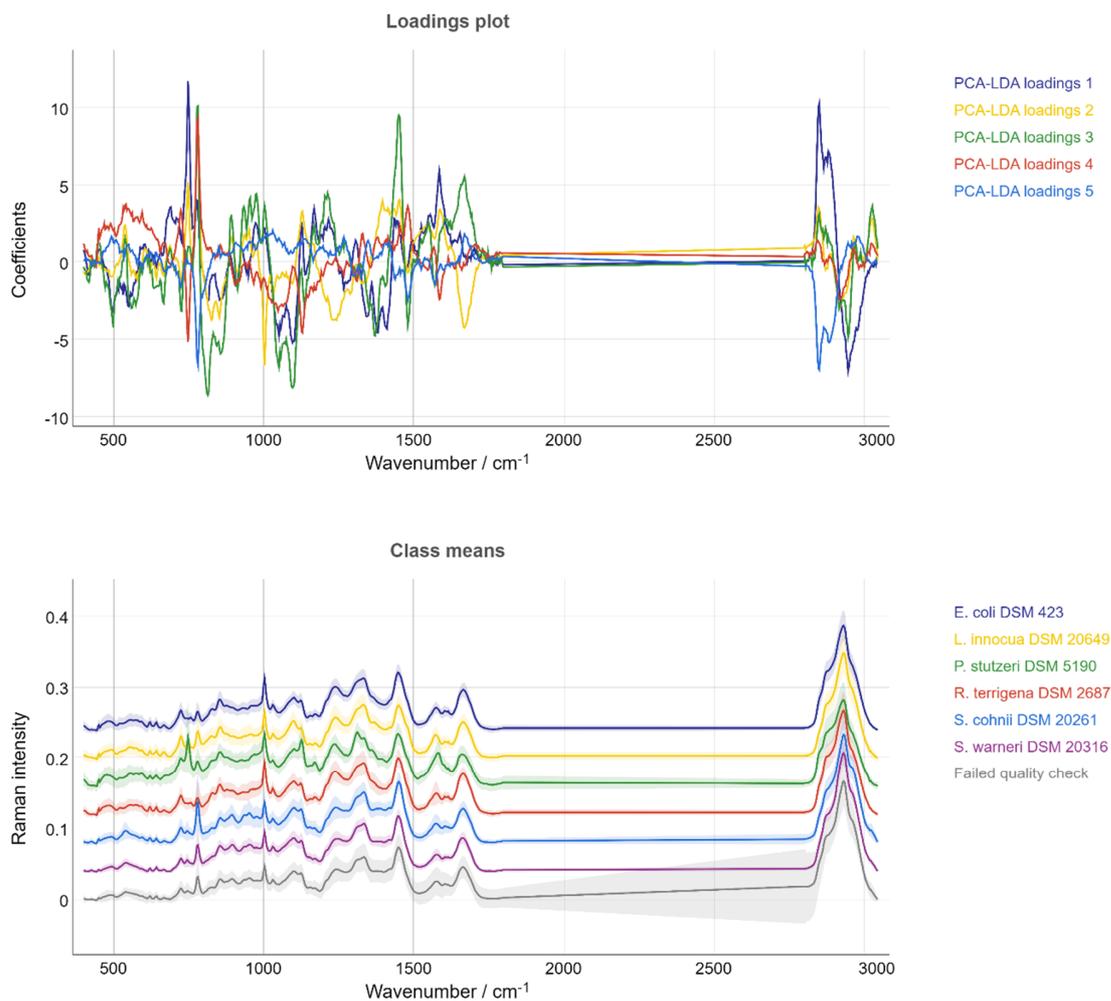

Figure 10 Model Loadings and mean spectra for the classes. PCA-LDA loadings plot helps to determine the peaks that contribute to the classification. Class means plot shows the preprocessed mean spectra per class after applying the quality filters. For better visualization, the mean spectra were stacked by adding an y-offset. Although the mean of the spectra that failed quality check does look similar to the other bacteria spectra, large standard deviation (shaded area) indicates much higher deviation of each individual spectrum from the average, which might be related to fluorescence background or noise.

Although it is good practice to apply 2-level cross-validation, it requires a higher number of batches (biological replicates) and may be quite time-consuming, especially for larger data sets. For the data sets with a large number of observations per sample, the possibility of averaging a selected number of spectra into a single spectrum is available. This feature might significantly speed-up the cross-validation routine, especially with computationally heavy models, such as SVM or RF.

Another feature that is quite useful for evaluating classification models based on data sets with many spectra per sample, is the voting option. By default, major voting is performed within groups defined by the combination of class and batch labels. Users may, however, customize the grouping based on columns provided in the metadata. Major voting makes it possible to obtain a single prediction for each group. These single predictions are then summarized using a confusion table.

## 5.5 Test prediction

There are two ways to create a test data prediction: using the model from the training data workflow or by importing a previously saved model. The second scenario corresponds to the GUI configuration of the lowest expertise level shown in Figure 11. In this configuration, there are no parameters to adjust and the user can only import test data and model (without the possibility to change it) and examine the results of the prediction.

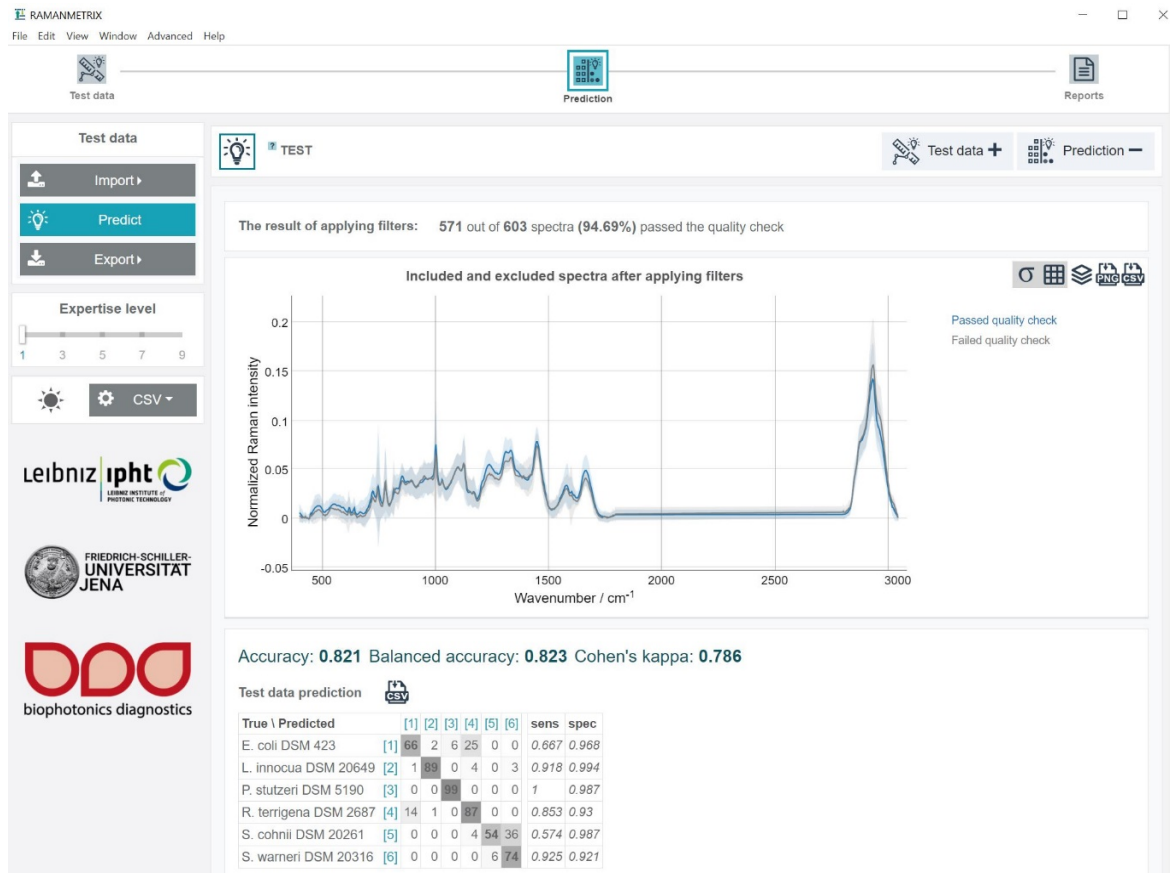

Figure 11 Test data prediction. The screenshot illustrates the **RAMAN**METRIX window view for the lowest expertise level. This limited functionality regime allows the user to only import test data and a prebuilt model and then generate a prediction.

After importing the test data, the mean spectrum over the dataset, mean spectra per class, and summary of the metadata will be displayed in the "Test data" tab. In this tab, the user can also start test data preprocessing and see the results. It is important to note here that the parameters utilized for the pretreatment and the preprocessing of the test data are the same parameters that were used to process the training data. If the user imports a previously saved model, parameters are already embedded in this file. The "Predict" button on the left menu will run test data preprocessing and construct a prediction in one-click. The results are presented in the "Prediction" tab as a table, mean spectra plot for each class, plot with spectra that passed and failed quality filters, and a scatterplot.

# 6 Exporting models and results

## 6.1 Saving of graphical results and printable reports

All the spectra, the scatter-matrix plot, the hierarchical tree, and quality filters can be saved as high-quality images in a PNG format. It is also possible to export the confusion table and mean spectra as CSV files. The corresponding buttons are located on the graph's toolbar, above the legend (Figure 3, Figure 6, Figure 11). By clicking on the "Report" icon in the top navigation menu, the user has the possibility to see all resulting plots and applied parameters for each processing step (for training or test dataset) within a single page. The summary report can be printed or saved in a PDF format at any stage in the program. The data processing report for the example training dataset can be found in the repository https://github.com/Bocklitz-Lab/supplementary_files/.

## 6.2 Export of models, parameters, and results

In the "Training data" block in the left menu bar (Figure 3) users can export parameters, model, and results. Parameters will be exported as a TXT file, which should not be manually corrected, but can be in the future imported using dropdown "Import" of the menu. The "Results" import option of "Training data" panel makes possible to save metadata, pretreated spectra, preprocessed spectra, and model scores for training data in a "data.zip" archive. Same functionality is available for the test data (Figure 11) in the respective panel.

After the model is created, the user has the opportunity to export it as a file, along with all used parameters. This feature is useful because the model can be shared with less experienced users who can import this model and build a prediction for a test dataset of their own. The model can be also reused by the third-party automated systems that use **RAMAN**METRIX as data processing backend.

# 7 Conclusions

Chemometric/machine learning based analysis of Raman spectra requires a complete preprocessing workflow including model construction, optimization, and model validation. Therefore, a reliable analysis of Raman spectra requires a specific set of skills and experiences. **RAMAN**METRIX can assist researchers that lack these skills and experience but want to use Raman spectra in their routine. As **RAMAN**METRIX provides a user-friendly GUI that makes it possible to apply such advanced chemometrics for Raman spectroscopic data analysis in an easy and intuitive way, it assists with the incorporation of Raman spectroscopic techniques into clinical routine. **RAMAN**METRIX has already shown its convenience in studies on bronchoalveolar bacteria with Raman spectroscopy[48], isotope probing of bacteria in visible and UV-Raman[49], noise sources in confocal Raman biosensing[50], detection of multi-resistant strains of *E. coli*[51], and dependency of bacterial phenotype from CO2[52].

The data processing and evaluation pipeline in the software is standardized according to the best practice guidance[53]. The fixed sequence of the data processing steps helps to avoid common mistakes and ensures reliability of the analysis. The calculation algorithms and the chemometric approaches, implemented in this software, including those developed in our group, have been validated in various types of biomedical Raman spectroscopic datasets [4,9,13,16,20,39,47].

Within this article, we demonstrated the **RAMAN**METRIX GUI and the data processing capabilities by constructing a classification model for differentiating bacterial species based on the corresponding measured Raman spectra. The full data processing workflow was presented, including data pretreatment, preprocessing, spectra quality evaluation, and predictive model construction. The model was validated by a 2-level cross-validation approach and the independent data set prediction. We showed how easy the adjustment of the parameters and the examination of the outputs can be in the Ramanmetrix GUI.

Utilization of the latest technologies in the field of software development made it possible to create a software tool that is available to users in two versions: a web application and a standalone program. The GUI has a modern and intuitive design. Due to the fact that the complexity of the GUI can be customized, the program can also be used by those who have limited or no experience in Raman data processing.

The separation between backend and frontend parts in the **RAMAN**METRIX architecture makes it possible to integrate **RAMAN**METRIX as a data processing backend, accessible through API by the measurement devices directly. By sending the raw measured spectra in JSON format to the **RAMAN**METRIX backend through API, the predictions from Raman spectra based on the preconstructed **RAMAN**METRIX model can be obtained in real-time.

In further development, we plan to improve the CNN feature extraction approach, expand the list of supported data formats, and add a few more baseline correction algorithms. We also plan to expand the user experience on the online platform by creating a personal account where users could save and share their Raman data analysis projects.

Instructions on how to access the online platform or install the desktop application can be found at https://docs.ramanmetrix.eu. The program is under continuous development and the authors will be happy to receive feedback.

## Competing interests

Authors have no conflict of interest to declare.

# Supplementary information:

# RAMANMETRIX: a delightful way to analyze Raman spectra

Darina Storozhuk[1,2 #], Oleg Ryabchykov[1,2 #], Juergen Popp[1,2 *], Thomas Bocklitz[1,2 *]

[1]Leibniz Institute of Photonic Technology Jena, Albert-Einstein-Str. 9, 07745 Jena, Germany,

[2]Institute for Physical Chemistry and Abbe Center of Photonics, Friedrich Schiller University, Helmholtzweg 4, 07743 Jena, Germany.

# equal contribution
* corresponding authors

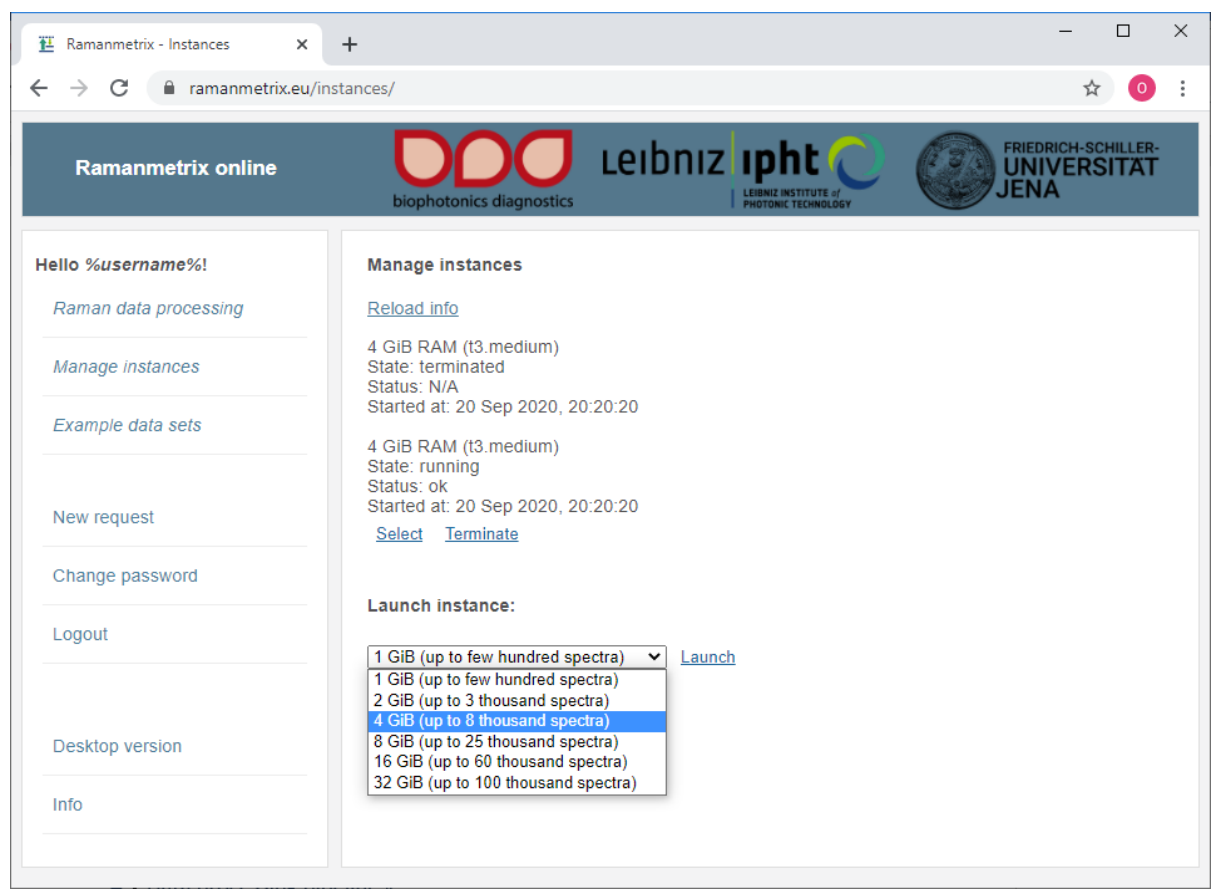

Figure S1 The user interface for launching a computational instance. Users have multiple options for instance RAM sizes, depending on the size of the data set. Besides launching new instances, users can switch between active instances or terminate them.

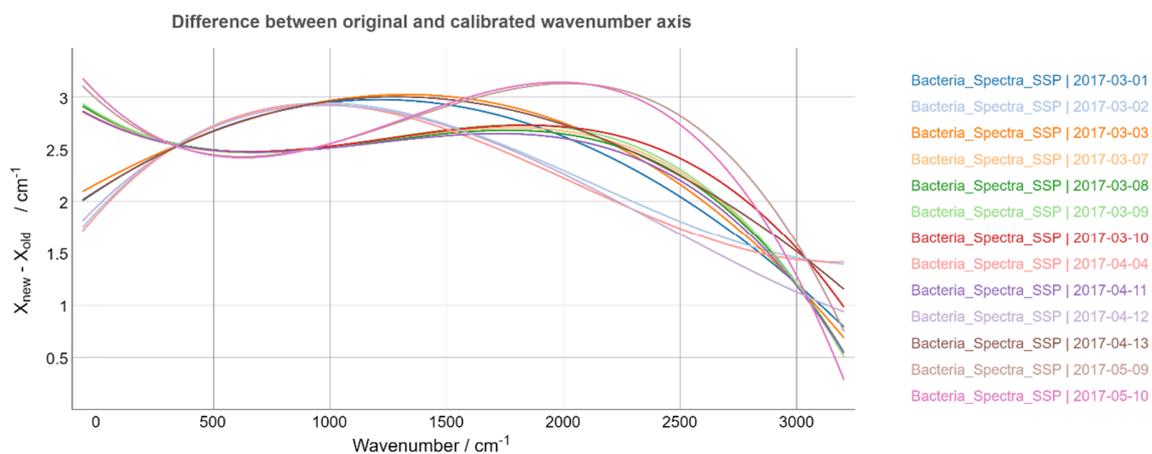

Figure S2 Difference between original and calibrated wavenumber axis. This graph illustrates the calibration functions for each measurement date. Unexpectedly large shift between the original wavenumber axis and the calibrated one indicate that the device calibration was largely off or that the calibration parameters need to be adjusted.

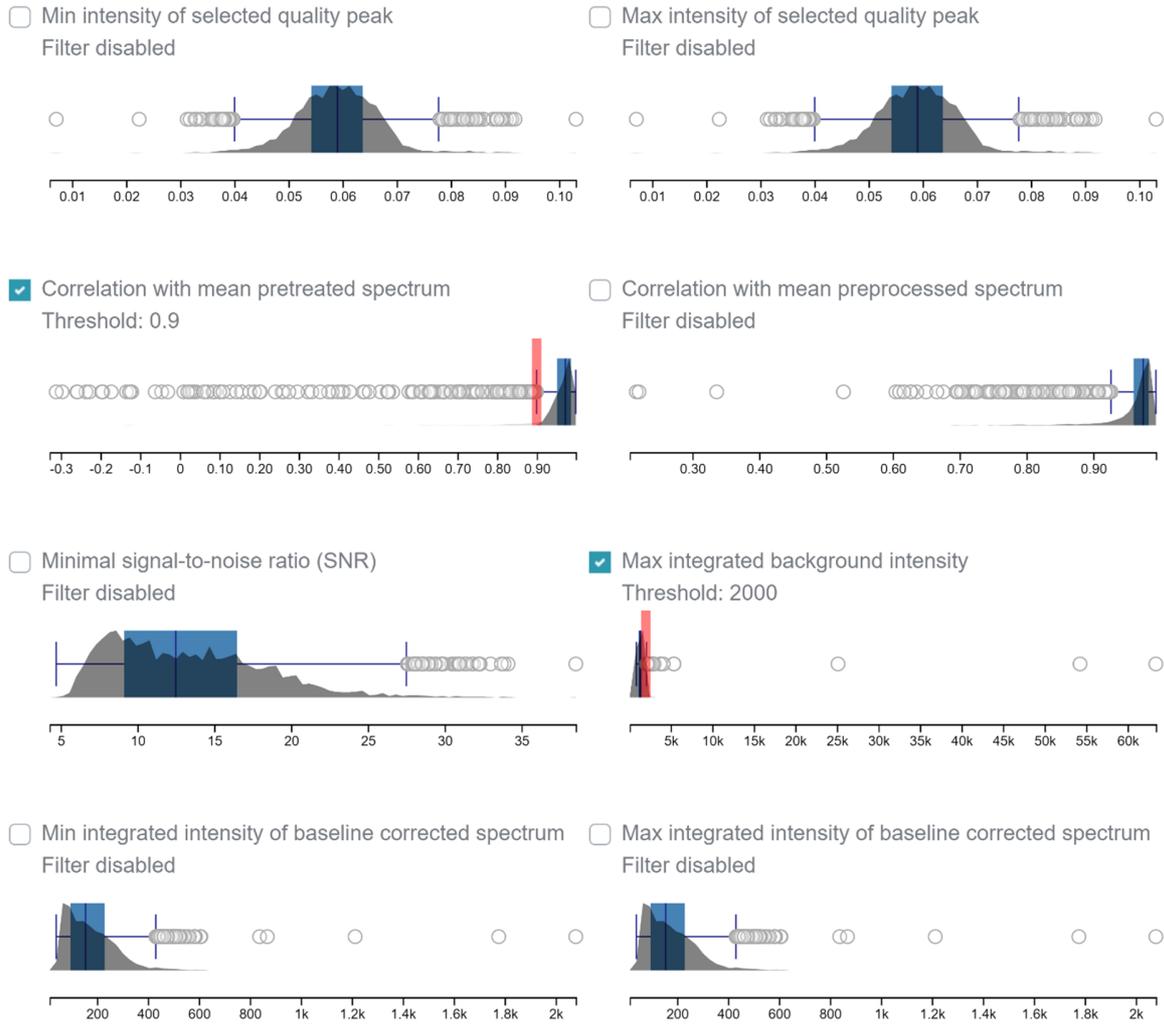

Figure S3 Quality filters. Different values related to the spectra quality can be used for filtering-out low-quality spectra. In the shown example, two out of 8 possible filters are enabled. As a result, only spectra that have correlation with the average pretreated (calibrated) spectrum above 0.9 and the background intensity below 2000 will be used in further analysis. The filtered-out spectra are marked as "Failed quality spectra" in subsequent mean spectra plots (Figure 10). The same thresholds are applied in the test data (Figure 11). In the test data set, average spectra from the training data are used to calculate the correlation.